\documentclass[10pt, twocolumn]{IEEEtran}

\usepackage{amsmath, mathrsfs,amsfonts}
\usepackage{amsfonts}
\usepackage{amssymb}
\usepackage{acronym}
\usepackage{graphicx}

\def\ignore#1{}

\def\bs{\boldsymbol}

\def\R{\mathbb{R}}

\def\foral{\textrm{for all} \ }

\newtheorem{definition}{Definition}
\newtheorem{theorem}{Theorem}
\newtheorem{lemma}{Lemma}

               {\begin{list}{}{\leftmargin#1\rightmargin#2\topsep#3}\item{}}%
               {\end{list}}

\begin{document}
\title{Information Theory vs. Queueing Theory for Resource Allocation in Multiple Access
Channels\thanks{This research was partially supported by the National Science Foundation under
grant DMI-0545910, and by DARPA ITMANET program.} \fontsize{11}{11}{  \\ \emph{(Invited Paper)}\\
}}


\author{Ali ParandehGheibi\thanks{A.\ ParandehGheibi is with the Laboratory for
Information and Decision Systems, Electrical Engineering and Computer Science Department,
Massachusetts Institute of Technology, Cambridge MA, 02139 (e-mail: parandeh@mit.edu)}, Muriel
M\'edard\thanks{ A.\ Ozdaglar and M.\ M\'edard are with the Laboratory for Information and Decision
Systems, Electrical Engineering and Computer Science Department, Massachusetts Institute of
Technology, Cambridge MA, 02139 (e-mails: asuman@mit.edu, medard@mit.edu)}, Asuman Ozdaglar, and
Atilla Eryilmaz\thanks{A.\ Eryilmaz is with the Electrical and Computer Engineering, Ohio State
University, OH, 43210 (e-mail: eryilmaz@ece.osu.edu)}}


\maketitle \thispagestyle{headings}

\begin{abstract}
We consider the problem of rate allocation in a fading Gaussian multiple-access channel with fixed
transmission powers. The goal is to maximize a general concave utility function of the expected
achieved rates of the users. There are different approaches to this problem in the literature. From
an information theoretic point of view, rates are allocated only by using the channel state
information. The queueing theory approach utilizes the global queue-length information for rate
allocation to guarantee throughput optimality as well as maximizing a utility function of the
rates. In this work, we make a connection between these two approaches by showing that the
information theoretic capacity region of a multiple-access channel and its stability region are
equivalent. Moreover, our numerical results show that a simple greedy policy which does not use the
queue-length information can outperform queue-length based policies in terms of convergence rate
and fairness.
\end{abstract}

\section{Introduction}
Dynamic allocation of communication resources such as bandwidth or transmission power is a central
issue in multiple access channels (MAC) in view of the time varying nature of the channel and
interference effects. Most of the existing literature on resource allocation in multiple access
channels focuses on specific communication schemes such as TDMA (time-division multiple access)
\cite{TDMA} and CDMA (code-division multiple access) \cite{CDMA1,CDMA3} systems. An exception is
the work by Tse \emph{et al.} \cite{Tse}, who introduced the notion of \emph{throughput capacity}
for the fading channel with Channel State Information (CSI) and studied dynamic rate allocation
policies with the goal of maximizing a linear utility function of rates over the throughput
capacity region.

Another important literature relevant to our work appears in the context of cross-layer design,
where joint scheduling-routing-flow control algorithms have been proposed and shown to achieve
utility maximization for concave utility functions while guaranteeing network stability (e.g.
\cite{linshr05, erysri05, neemodli05, sto05}). The common idea behind these schemes is to use
properly maintained queues to make dynamic decisions about new packet transmission as well as rate
allocation.

Some of these works (\cite{erysri05, neemodli05}) explicitly address the fading channel conditions,
and show that the associated policies can achieve rates arbitrarily close to the optimal based on a
design parameter choice. Requiring global queue-length information may impose much more overhead on
the system than channel state information, because for a slow fading channel the sampling rate for
channel estimation is significantly smaller than the data arrival rate which dictates the necessary
sampling rate for obtaining the queue-length information.

In this paper, we consider the problem of rate allocation in a multiple access channel with perfect
CSI. We assume that the transmitters do not have power control feature, and no prior knowledge of
channel statistics is available. We first relates the information theory and queueing theory
approaches by showing the equivalence between the information theoretic capacity region of a
multiple-access channel and its stability region. Hence, we conclude that the policies assigning
rates form the capacity region of a multiple-access channel are indeed throughput optimal.

We consider the utility maximization framework with a general concave utility function of the rates
to model different performance metrics and fairness criteria (cf.\ Shenker \cite{She95}, Srikant
\cite{Srikant}). We present an efficient greedy rate allocation policy \cite{MAC_Wiopt} which is in
general suboptimal but does not require queue-length information, and compare its performance with
the queue-length based policy by Eryilmaz and Srikant \cite{erysri05} for different scenarios with
limited communication duration.


The remainder of this paper is organized as follows: In Section II, we introduce the model and
describe the capacity region of a multiple-access channel. In Section III, we relate the
information theory and queueing theory approaches to resource allocation. In Section IV, we present
a dynamic rate allocation policy designed by each of theses approaches. In Section V, we provide
the simulation results to compare the performance of the presented policies under different
scenarios. Finally, we give our concluding remarks in Section VI.

\section{System Model}
We consider $M$ users sharing the same media to communicate to a single receiver. We model the
channel as a Gaussian multiple access channel with flat fading effects
\begin{equation}\label{fading_model}
    Y(t) = \sum_{i=1}^M  \sqrt{H_i(t)} X_i(t) + Z(t),
\end{equation}
where $X_i(t)$ are the transmitted waveform with average power $P_i$, $H_i(t)$ is the channel gain
corresponding to the \textit{i}-th user and $Z(t)$ is white Gaussian noise with variance $N_0$. We
assume that the channel gains are known to all users and the receiver \footnote{This is assumption
is satisfied in practice when the receiver measures the channels and feeds back the channel
information to the users.}. Throughout this work we assume that the transmission powers are fixed
and no prior knowledge of channel statistics is available.

First, consider the non-fading case where the channel gains are fixed. The capacity region of the
Gaussian multiple-access channel with no power control is described as follows \cite{cover}
\begin{eqnarray}\label{Cg}
    C_g(\bs P, \bs h) &=& \bigg\{ \bs R \in \mathbb{R}^M_+: \sum_{i \in S} R_i \leq  C\Big(\sum_{i \in S} h_i P_i,
    N_0\Big), \nonumber \\
     &&\quad \qquad  \textrm{for all}\  S \subseteq \mathcal M = \{1,\ldots, M\} \bigg\},
\end{eqnarray}
where $P_i$ and $R_i$ are the \emph{i}-th transmitter's power and rate, respectively. $C(P,N)$
denotes Shannon's formula for the capacity of an AWGN channel given by
\begin{equation}\label{C_AWGN}
    C(P,N) = \frac{1}{2}\log(1+\frac{P}{N}) \quad \textrm{nats}.
\end{equation}

For a multiple-access channel with fading, but fixed transmission powers $P_i$, the
\emph{throughput} capacity region is given by averaging the instantaneous capacity regions with
respect to the fading process \cite{Shamai},
\begin{eqnarray}\label{Ca}
    C_a(\bs P) &=& \bigg\{ \bs R \in \mathbb{R}^M_+: \sum_{i \in S} R_i
    \leq \mathbb{E}_{\bs H} \bigg[ C\Big(\sum_{i \in S} H_i P_i, N_0\Big) \bigg], \nonumber \\
    && \quad \qquad \qquad \qquad  \textrm{for all} \  S  \subseteq \{1, \ldots, M\} \bigg\},
\end{eqnarray}
where $\mathbb{E}_{\bs H}$ denotes the expectation with respect to $\bs H$, that is a random vector
with the stationary distribution of the fading process. Under the utility maximization framework,
the goal is to find optimal rate allocation policy with respect to the utility function,
$u(\cdot)$, as defined in the following.
\begin{definition}\label{optimal_policy}
[Optimal Policy] The optimal rate allocation policy denoted by $\mathcal{R}^*:\R^M \rightarrow
\R^M$ is a mapping that satisfies $\mathcal{R}^*(\bs H) \in C_g\big(\bs P,\bs H\big)$ for all $\bs
H$, and
\begin{eqnarray}\label{RAC}
\mathbb{E}_{\boldsymbol{H}} [\mathcal{R}^*(\bs H)] = \bs R^* \in& \textrm{argmax}& \quad u(\bs R)
\nonumber \\
&\textrm{subject to}& \quad \bs R \in  C_a({\bs P}).
\end{eqnarray}
\end{definition}

\section{Stability Region and Information Theoretic Capacity Region}\label{Q_sec}
In this part, first we make a connection between information theory and queueing theory approaches
by showing the information theoretic capacity region of a multiple access channel and its stability
region coincide. The stability region is defined as the closure of the set of arrival rates for
which there is a service scheduling policy stabilizing the queues, i.e., $\limsup_{t \rightarrow
\infty} \mathbb E[Q_i(t)] < \infty$, where $Q_i(t)$ is the queue length of user $i$. This approach
is similar to the work of M\'edard \emph{et al.} \cite{capacity_muriel} in which stability of a
time-slotted Aloha system is established, regardless of the burstiness of the traffic. Let us first
introduce a useful lemma known as \emph{$T$-slot Lyapunov drift} that is widely used in stability
analysis of queueing systems.

\begin{lemma}\label{Lyapunov_lemma}
(\emph{$T$-slot Lyapunov drift}\cite{RAC_book}) Let $\bs Q(t) \in \R^M$ denote the queue-lengths in
a network with $M$ users. If there exists a positive integer $T$ such that $\mathbb E[\bs Q(\tau)]
< \infty$ for $\tau \in \{1, \ldots, T-1\}$, and if there are positive scalars $\epsilon$ and $B$
such that for all time slots $t_0$ we have:
\begin{equation}
\mathbb E \Big[ V(\bs Q(t_0+T)) - V(\bs Q(t_0)) \mid \bs Q(t_0)\Big] \leq B - \epsilon \sum_{i=1}^M
Q_i(t_0), \nonumber
\end{equation}
for some Lyapunov function $V: \R^M \rightarrow \R$, then the network is strongly stable, i.e., the
queue-lengths satisfy:
$$\limsup_{t \rightarrow \infty} \frac{1}{t}\sum_{\tau =0}^{t-1} \sum_{i=1}^M \mathbb[Q_i(\tau)] \leq \frac{B}{\epsilon}  .$$
\end{lemma}

Next, we present a new result which establishes the equivalence between the capacity region and
stability region of a Gaussian MAC.
\begin{theorem}\label{stability_capacity}
Consider a $M$-user Gaussian multiple access channel. Let $A_i(t)$ be the arrival rate of the
$i$-th user at time slot $t$, where $\mathbb E[A_i(t)] = \lambda_i$ and $\mathbb E[A_i^2(t)] <
\infty$. Then the capacity region of the channel coincides with its stability region.
\end{theorem}
\begin{proof}
Let $\Lambda$ and $C$ denote the stability region and information theoretic capacity region of the
channel, respectively. First, we show that $C \subset \Lambda$ by presenting a simple achievable
scheme based on achievability proof of MAC capacity region which guarantees stability of the queues
for any achievable rate tuple.

Assume for the expected arrival rate $\bs \lambda$ we have $\bs \lambda \in \textrm{int}(C)$. Then,
there exists a rate tuple $\bs R \in \textrm{int}(C)$ such that $R_i = \lambda_i + \epsilon$, for
some $\epsilon > 0$. Since $\bs R$ lies in the interior of $C$, it is achievable, i.e., for every
$P_e >0$ there exists large enough $N$ such that user $i$ can transmit $N R_i$ bits of information
by $N$ channel use with probability of error smaller than $P_e$. For the given rate tuple $\bs R
\in \textrm{int}(C)$, choose $n$ large enough such that
\begin{equation}\label{error_prob}
 P_e \leq \frac{\epsilon}{2(\max_i\lambda_i + \epsilon)}.
\end{equation}

Using the Foster's criterion for stability, we show the following scheduling algorithm stabilizes
the queues. User $i$ starts transmitting only if  $ Q_i(t) - n R_i < 0$, where $Q_i(t)$ denotes the
queue-length of user $i$ at time $t$. In this case user $i$ chooses a codeword from its codebook of
size $2^{n R_i}$ and transmits it over the channel by $n$ channel use.

Note that in this scheme users transmit asynchronously. However, the information theoretic capacity
region remains the same \cite{async_MAC}. The $i$-th user's queue dynamics are given by
\begin{equation}\label{Q_dynamic}
    Q_i(t+n) = \left\{ \begin{array}{ll}
                   Q_i(t) + \sum_{\tau = t}^{t+n-1} A_i(\tau), &\!\!\!\!\!\!\!\! Q_i(t) < nR_i\\
                   Q_i(t) - n R_i S_i(t) + \sum_{\tau = t}^{t+n-1} A_i(\tau), & \textrm{otherwise.} \\
                   \end{array} \right.
\end{equation}
where $Q_i(t)$ denotes the queue-length of user $i$, and $S_i(t)$ are Bernoulli random variables
that take value 0 with probability $P_e$. Let $V(\bs Q) = \sum_{i=1}^M Q_i^2$ be the Lyapunov
function, and define the $n$-slot drift as follows:
\begin{eqnarray}\label{Lyapunov_drift}
    \Delta_n \big(\bs Q(t)\big) &=&  \mathbb{E}\Big[ V\big(\bs Q(t+n)\big) - V\big(\bs Q(t)\big) \mid \bs
    Q(t)\Big] \\
\label{drift_i}    &=& \sum_{i=1}^M  \mathbb{E}\Big[Q_i^2(t+n) -Q_i^2(t) \mid \bs Q(t)\Big].
\end{eqnarray}

Given the queue dynamics in (\ref{Q_dynamic}), we can bound each term in (\ref{drift_i}) as
follows. For the case that $Q_i(t) \geq nR_i = n(\lambda_i +\epsilon)$, we have
\begin{eqnarray}\label{drift_bd1}
&&    \mathbb{E}\Big[Q_i^2(t+n) -Q_i^2(t) \mid \bs Q(t)\Big] \nonumber \\
&& \quad   = \mathbb{E}\Big[\big(Q_i(t+n) -Q_i(t)\big)\big(Q_i(t+n) +Q_i(t)\big) \mid \bs Q(t)\Big] \nonumber \\
&& \quad    = \mathbb{E}\Big[2Q_i(t) \big( \sum_{\tau = t}^{t+n-1} A_i(\tau) - n R_i S_i(t) \big)  \mid \bs  Q(t)\Big] \nonumber   \\
&& \qquad  +\ \mathbb{E}\Big[\big( \sum_{\tau = t}^{t+n-1} A_i(\tau) - n R_i S_i(t) \big)^2 \mid \bs Q(t)\Big]  \nonumber \\
&& \quad = 2Q_i(t) \Big(n\lambda_i -n(\lambda_i+\epsilon)(1-P_e) \Big) \nonumber \\
&& \qquad  + \ n\mathbb \mathbb{E}[A_i^2(t)] + n(n-1)\lambda_i^2 + n^2 R_i^2 (1-P_e)  \nonumber \\
&& \qquad  - \  2n^2 \lambda_i R_i (1-P_e) \nonumber \\
&& \quad = 2n Q_i(t)\big(P_e(\lambda_i +\epsilon) - \epsilon\big) +B'_i \nonumber \\
&& \quad \leq -(n \epsilon) Q_i(t) + B'_i =  -\epsilon' Q_i(t) + B'_i, \nonumber
\end{eqnarray}
where $B'_i \leq n \mathbb{E}[A_i^2(t)] + n^2(P_e \lambda_i + \epsilon R_i) <\infty$, and the
inequality follows from (\ref{error_prob}).

Now consider the case that $Q_i(t) < nR_i$, where $n$ is  a fixed finite number. Since the
queue-length is bounded at time $t$, it will remain bounded at time $t+n$, even if no successful
transmission occurs between $t$ and $t+n$. Thus, we can write a negative drift for $Q_i$. In
particular
\begin{eqnarray}\label{drift_bd2}
&&  \!\!\!\!\!\!\!\!  \mathbb{E}\Big[Q_i^2(t+n) -Q_i^2(t) \mid \bs Q(t)\Big] \nonumber \\
&& \!\!\!\! =  \mathbb{E}\Big[2Q_i(t) \big( \sum_{\tau = t}^{t+n-1} A_i(\tau)\big)  \mid \bs  Q\Big] +\ \mathbb{E}\Big[\big( \sum_{\tau = t}^{t+n-1} A_i(\tau) \big)^2 \mid \bs  Q\Big]    \nonumber \\
&& \!\!\!\!   = 2Q_i(t) \Big(n\lambda_i -\frac{n\epsilon}{2} +\frac{n\epsilon}{2}\Big) + n\mathbb E[A_i^2(t)] + n(n-1)\lambda_i^2  \nonumber \\
&& \!\!\!\!  = -(n \epsilon) Q_i(t) + n^2R_i(2\lambda_i + \epsilon) + n\mathbb E[A_i^2(t)] + n(n-1)\lambda_i^2  \nonumber \\
&& \!\!\!\!    \leq - \epsilon' Q_i(t) + B''_i, \nonumber
\end{eqnarray}
where $B''_i =n^2R_i(2\lambda_i + \epsilon) + n\mathbb E[A_i^2(t)] + n(n-1)\lambda_i^2  <\infty$.

The total Lyapunov drift can be written as
\begin{eqnarray}\label{Lyapunov_drift2}
    \Delta_n \big(\bs Q(t)\big) &=& \sum_{i=1}^M \mathbb E\Big[Q_i^2(t+n) -Q_i^2(t) \mid \bs Q(t)\Big]
    \nonumber \\
    &\leq& - \epsilon' \sum_{i=1}^M Q_i(t) + B,
\end{eqnarray}
where $B= \sum_{i=1}^M \max\{B'_i,B''_i\} < \infty$.

Therefore, by Lemma \ref{Lyapunov_lemma}, the network is stable for any arrival rate $\bs \lambda$
in the interior of the information theoretic capacity region. Hence, $C \subset \Gamma$.

For the converse part, suppose $\bs \lambda \notin C$. We show that no scheduling policy can
stabilize the queues. Since $\bs \lambda \notin C$, there exists some $S \subseteq \mathcal M$ such
that $$\sum_{i \in S} \lambda_i \geq C(\sum_{i \in S}P_i,N_0) + \epsilon,$$ for some $\epsilon >
0$. It follows from Fano's inequality \cite{cover} that transmission rate tuple $\bs R$ per channel
use and the probability of error, $P_e$, satisfy the following
\begin{equation}\label{Fano_MAC}
    \sum_{i \in S}R_i \leq C(\sum_{i \in S}P_i,N_0) + P_e \sum_{i \in S}R_i.
\end{equation}

Therefore, by the Strong Law of Large Numbers we have
\begin{eqnarray}
  \sum_{i \in S}Q_i(t) &=& \sum_{i \in S}\sum_{\tau=0}^{t-1} A_i(\tau) - t \sum_{i \in S}R_i(1-P_e)
  \nonumber \\
  &=& t\sum_{i \in S} \lambda_i -  t(1-P_e) \sum_{i \in S}R_i \nonumber \\
  &\geq& t\bigg(C(\sum_{i \in S}P_i,N_0) +\epsilon \bigg) - t C(\sum_{i \in S}P_i,N_0) = t\epsilon,
  \nonumber
\end{eqnarray}
where the inequality directly follows from (\ref{Fano_MAC}). By taking the limit as $t$ goes to
infinity, we conclude that not all the queues in the network can be stabilized. This establishes
the converse and hence, equivalence of the stability region and information theoretic region of the
multiple access channel.
\end{proof}

The proof of Theorem \ref{stability_capacity} simply generalizes for the case of fading multiple
access channel using the achievability proof of fading MAC and by taking the expectations with
respect to the fading process. Next, we present dynamic rate allocation policies designed from
information theory and queueing theory point of view.

\section{Dynamic Rate Allocation Policies}\label{policy_sec}
In this section, we first present a greedy rate allocation policy that only uses the instantaneous
channel state information to allocate rates to different users.
\begin{definition}\label{greedy_policy}
[Greedy Policy] A \emph{greedy} rate allocation policy, denoted by $\mathcal{\bar{R}}$, is given by
\begin{eqnarray}\label{RAC_greedy}
\mathcal{\bar{R}}(\bs H) = & \textrm{argmax}& \quad u(\bs R)
\nonumber \\
&\textrm{subject to}& \quad \bs R \in  C_g({\bs P}, \bs H)
\end{eqnarray}
i.e., for each channel state, the greedy policy chooses the rate vector that maximizes the utility
function over the corresponding capacity region.
\end{definition}
This policy can be implemented efficiently by exploiting polymatroid structure of MAC capacity
region leading to a gradient projection method with approximate projection \cite{MAC_ITA}. Note
that the greedy policy is not necessarily optimal for general concave utility functions. However,
it can be shown \cite{MAC_Wiopt} that under some mild technical assumptions, the performance
difference between the greedy and the optimal policy (cf. Definition \ref{optimal_policy}) is
bounded. This bound goes to zero as the utility tends to a linear function or the channel
variations vanish.

Now, we present a rate allocation policy for fading multiple access channel by Eryilmaz and Srikant
\cite{erysri05}. This policy explicitly uses queue-length information in order to guarantee
stability of the network and also captures some fairness issues by considering the following
utility function:
\begin{equation}\label{a_fair_util}
    u(\bs R) = \sum_i{w_i f_\alpha(R_i)},
\end{equation}
where $f_\alpha(\cdot)$ denotes an $\alpha$-fair function given by
\begin{equation}\label{a_fair}
    f_{\alpha}(x) = \left \{ \begin{array}{ll}
                   \frac{x^{1-\alpha}}{1-\alpha}, & \textrm{$\alpha \neq 1$}\\
                   \log(x), & \textrm{$\alpha = 1.$}
                   \end{array} \right.
\end{equation}

\begin{figure}
  \centering
  \includegraphics[width=.32\textwidth]{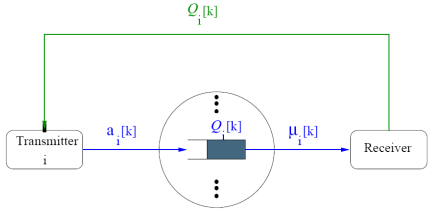}\\
  \caption{Structure of the $i$-th transmitter and the receiver for the queue-length-based policy \cite{erysri05}.}\label{queue_fig}
\end{figure}

As illustrated in Figure \ref{queue_fig}, $Q_i(t)$ is the queue-length of the $i$-th user. The
evolution of the $i$-th queue is given by
$$Q_i(t+1) = \Big(Q_i(t) + a_i(t) - \mu_i(t)\Big)^+,$$
where $a_i(t)$ and $\mu_i(t)$ denote the arrivals and service of queue $i$, respectively. At time
slot $t$, the scheduler chooses the service rate vector $\bs \mu (t)$ based on a max-weight policy,
i.e.,
\begin{eqnarray}\label{MW_scheduler}
   \bs \mu(t) = &\textrm{argmax}& \sum_{i=1}^M Q_i(t)R_i
\nonumber \\
&\textrm{subject to}& \quad \bs R \in  C_g({\bs P}, \bs H(t)).
\end{eqnarray}

The congestion controller proposed in \cite{erysri05} leads to a fair allocation of the rates for a
given $\alpha$-fair utility function. In particular, the data generation rate for the $i$-th
transmitter, denoted by $a_i(t)$ is a random variable satisfying the following conditions:
\begin{eqnarray}
  \mathbb E\big[a_i(t)\ | Q_i(t)\big] &=& \min \bigg\{K\Big(\frac{w_i}{Q_i(t)}\Big)^{\frac{1}{\alpha}}, D  \bigg\}, \nonumber \\
    \mathbb E\big[a_i^2(t)\ | Q_i(t)\big] &\leq& U < \infty, \quad \foral Q_i(t),
\end{eqnarray}
where $\alpha$, $D$ and $U$ are positive constants. The following theorem from \cite{erysri05}
states that the parametric rate allocation policy presented above is asymptotically optimal as $K$
grows to infinity.

\begin{theorem}
Let $\bs R^*$ be the optimal solution of the optimization problem (\ref{RAC}). The mean stationary
service rate vector converges to $\bs R^*$ as $K$ increases, i.e.,
$$|\bar{\mu}_i - R_i^*| \leq B(K)^{-\alpha/2},$$
where $B$ is a finite positive constant and $\bar{\bs \mu}$ is the mean service rate provided by
the rate allocation algorithm under the stationary distribution of the Markov chain. Moreover, for
any initial queue-length vector $\bs Q(0)$, we have
$$\lim_{T \rightarrow \infty} \frac{1}{T} \sum_{t=0}^{T-1} \min \bigg\{ K\Big(\frac{w_i}{Q_i(t)}\Big)^{\frac{1}{\alpha}}, D\bigg\} = \bar{\bs \mu}, \quad \textrm{almost surely.}$$
\end{theorem}

The proof and further details are given in \cite{erysri05}. In  the following, we shall study the
performance of this policy further using numerical results in the following section.

\section{Simulation Results and Discussion}\label{simulation_chap}
In this section, we provide simulation results to compare performance of the rate allocation
policies presented in Section \ref{policy_sec}. For this simulations, we make the reasonable
assumption that the channel state processes are generated by independent identical finite state
Markov chains.

We consider two different scenarios to compare the performance of the greedy policy with the
queue-based rate allocation policy by Eryilmaz and Srikant \cite{erysri05}. This policy,
parameterized by some parameter $K$, uses queue length information to allocated the rates
arbitrarily close to the optimal policy by choosing $K$ large enough. The greedy policy as defined
in Definition \ref{greedy_policy} only uses the channel state information and does not require
queue-length information for dynamic rate allocation.

%

\begin{figure}
  \centering
  \includegraphics[width=.32\textwidth]{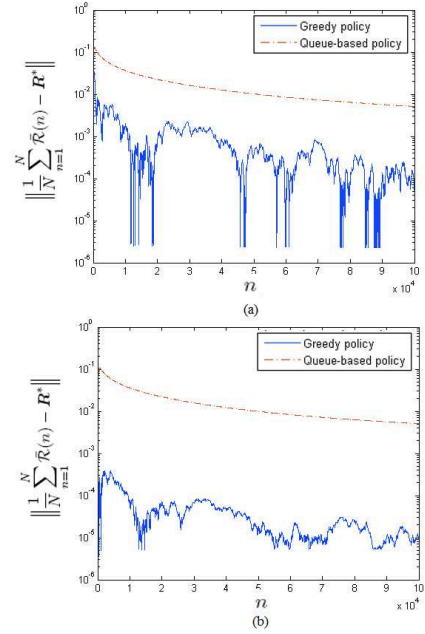}
 \caption{Performance comparison of the greedy and the queue-based policies for a communication session with
 limited duration, for (a) $\frac{\sigma_H}{\bar H} = 1.22$ (b) $\frac{\sigma_H}{\bar H} =
 0.13$.}\label{greedy_Q_fig}
\end{figure}

In the first scenario, we compare the average achieved rate of the policies for a communication
session with limited duration. Figure \ref{greedy_Q_fig}(a) depicts the distance between empirical
average achieved rate of greedy or queue-length based policy and $\bs R^*$, the maximizer of the
utility function over throughput region. In this case, the utility function is given by
(\ref{a_fair_util}) with $\alpha = 2$ and $w_1 = 1.5w_2 = 1.5$, and the corresponding optimal
solution is $\bs R^* = (0.60, 0.49)$. As observed in Figure \ref{greedy_Q_fig}(a), the greedy
policy outperforms the queue-length based policy for limited duration of the communication session.
It is worth noting that there is a tradeoff in choosing the parameter $K$ of the queue-length based
policy. In order to guarantee achieving close to optimal rates by queue-based policy, the parameter
$K$ should be chosen large which results in large expected queue length and lower convergence rate.
On the other hand, if $K$ takes a small value to improve the convergence rate, the expected
achieved rate of the queue based policy lies in a larger neighborhood of the $\bs R^*$.

Furthermore, the performance of the greedy policy improves by decreasing the channel variations.
Figure \ref{greedy_Q_fig}(b) demonstrates the improvement in performance of the greedy policy when
the total channel variations, $\frac{\sigma_H}{\bar{H}}$, has decreased from 1.22 to 0.13. We also
observe in Figure \ref{greedy_Q_fig}(b) that the queue-length based policy does not improve by
decreasing channel variations.

\begin{figure}
  \centering
  \includegraphics[width=.3\textwidth]{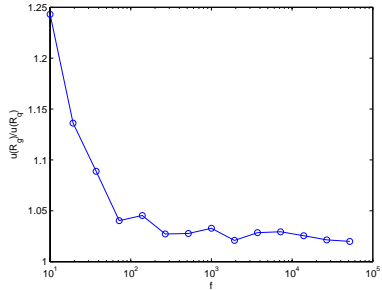}\\
  \caption{Performance comparison of greedy and queue-based policies for file upload scenario with
  respect to file size $f = f_1 = f_2$. $R_g$ and $R_q$ are expected upload rate of the greedy and
  the queue-length based policy, respectively.}\label{greedy_Q_upload_fig}
\end{figure}
Second, we consider a file upload scenario where each user transmits a file with finite size to the
base station in a rateless manner. Let $\mathcal{T}_i$ be the $i$-th user's completion time of the
file upload session for a file of size $f_i$.  Define the average upload rate for the $i$-th user
as $\frac{f_i}{\mathcal{T}_i}$. We can measure the performance of each policy by evaluating the
utility function at the average upload rate. Figure \ref{greedy_Q_upload_fig} illustrates the
utility difference of the greedy and the queue-based policy for different file sizes. We observe
that for small file sizes the greedy policy outperforms the queue-based policy significantly, and
this difference decrease by increasing the file size. We can interpret this behavior as follows.
The files are first buffered into the queues based on the queue lengths and the weighted
$\alpha$-fair utility, while the queues are emptied by a max-weight scheduler. Once the files are
all buffered in the queues, the queues empty with the same rate which is not fair because it does
not give any priority to the users based on their utility. For larger file sizes, the duration for
which the queue is non-empty but the file is totally buffered in the queue, is negligible compared
to total transmission time, and the average upload rate converges to a nearly optimal rate.

\section{Conclusion}
We addressed the problem of rate allocation in a fading multiple access channel with no power
control and prior knowledge of channel statistics, and considered the utility maximization
framework for a general concave utility function of the rates. We made a connection between the
information theory and queueing theory approaches to this problem by showing the equivalence
relation between the capacity region of a multiple access channel and its stability region.

We also presented dynamic rate allocation policies designed by each approach and compared their
performance using the simulation results for a communication scenario with limited duration or
limited file size. The numerical results show that a simple greedy policy that does not use the
queue-length information can outperform the queue-length based policies in terms of convergence
rate and fairness.

\bibliographystyle{unsrt}
\bibliography{MAC}

\end{document}